# Physics Faculty and Empathy in Academic Spaces

Rachel Merrill, Yale University, rachel.merrill@yale.edu
Alia Hamdan, Ash Bista, Scott Franklin
ajhcos@rit.edu, ab7878@rit.edu, svfsp@rit.edu
Rochester Institute of Technology

**Abstract:** The ability to emotionally or intellectually understand another person's thoughts and feelings — empathy — can foster critical connections that facilitate learning and collaboration. We present a case study of physics faculty that examines their experiences empathizing with students, both in and outside of the classroom. We expand on frameworks for understanding the empathy process by identifying key mediating factors, and note various barriers that faculty express as preventing them from taking empathetic action. Our analysis unpacks the mechanisms of communication and contextual information, which play key roles in the empathetic process, with implications for programs that rely on empathy to develop more inclusive STEM academic spaces.

## Introduction

Empathy is a multifaceted construct defined as "the ability and tendency to share and understand others' internal state" (Zaki & Ochsner, 2012) and is foundational to an effective learning space. Empathy builds strong relationships (Lunn et al., 2022), and instructors displaying higher empathy towards students have improved impacts on student achievement (Postolache, 2020) and receive more respect from these students (Feshbach & Feshbach, 2009). Faculty recognize this and develop their capacity for empathy as they learn more about their students and the issues they face (Lunn et al., 2022). Research on empathy in the classroom is multidisciplinary, with growing interest in medicine, engineering and other disciplines (Eklund & Meranius, 2021).

Here we investigate mediating factors between *noticing* inequities, *empathizing* with a student or colleague, and deciding to take *empathetic action*. Dancy and Hodari (2023) found progressive, well-intentioned, and educated white male physicists often struggle with interpreting others' affective states in response to racist or sexist incidents. This struggle, coupled with a failure to identify the larger implications to victims of racism or sexism, led to inaction or deflection of responsibility. Here we identify a struggle to communicate meaningfully as a primary marker of inaction or ineffective empathetic action. This lack of communication has been shown previously to negatively impact a student's decision to continue in the discipline (Lillis, 2011; Park et al., 2020), with attrition particularly impacting students from historically marginalized backgrounds. To better understand the root causes of action and inaction, we investigate *how physics faculty experience and apply empathy when interacting with students*, with consequences for the accepted model of "empathetic pathway(s)" (Yu & Chou, 2018).

## Theory: Empathetic pathway(s) mediated by communication

Yu and Chou (2018) proposed a framework for how empathy can lead to action. The process begins with an individual's attention being drawn to the other party (*noticing*), which triggers empathy and can then lead to empathetic action. *Noticing*, which triggers empathy, is not automatic, and multiple studies have found that people in privileged positions, in particular white men, struggle to identify acts of sexism, racism and ableism (Dancy et al., 2020; Rodin et al., 1990; Swim et al., 2001). Two distinctly different types of empathy have been identified: *cognitive* and *affective/emotional*. Cognitive empathy is one's ability to intellectually understand another person's perspective or emotions (Gladstein, 1983), while affective or emotional empathy entails emotionally relating to the other person's experience or feelings (Reniers et al., 2011). Yu and Chou (2018) provide a neurological framework for these different processes, with emotional empathy being a reflexive neurological response, occurring almost instantaneously, and cognitive empathy requiring a conscious intellectual effort. Dancy (personal communication, October 4, 2023) noted the preponderance of cognitive empathy, and the relative lack of emotional empathy, in how their participants discussed issues of racism and sexism and speculated that this explains their inaction. The empathetic process is illustrated in Fig. 1.

**Figure 1**
*Empathetic pathway as illustrated from the accumulation of past literature (e.g.Yu and Chou, 2018), with key concepts (boxes) connected by mediating factors (arrows).*

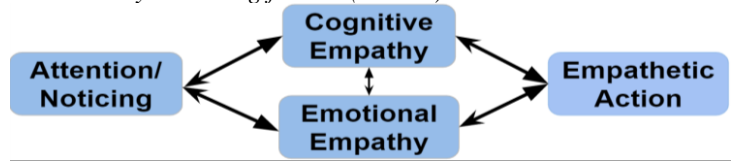

Eklund and Meranius (2021) define four traits required for an individual to enact empathy: understand[ing], feel[ing], shar[ing] another person's world, and self-differentiat[ing] oneself from another. While the framework in Fig. 1 identifies pathways toward empathy development and subsequent actions, it overlooks the essential aspect of empathy evolving between individuals. Models of the communication process (Betts, 2009) include linear (sender transmits to passive receiver), interactive (dynamic communication loop) and transactional (collaborative construction that includes nonverbal and cultural messaging). Viewing empathy as a communication process emphasizes active participation and understanding beyond verbal exchanges (Kunyk & Olson, 2001). Here we frame empathy as a collaborative construct that emerges through shared communication and understanding between individuals, building on prior models to understand how faculty use communication to experience and apply empathy toward students.

## Methods

Eight physics faculty (seven tenure/tenure-track, one non-tenured teaching position) from a private research university in the northeastern United States were interviewed using a semi-structured protocol. Each participant was interviewed twice: first an introductory interview to identify how and when faculty interacted with others in academic spaces, and a second follow-up interview to explore specific episodes mentioned in the first interview. Follow-up interviews revisited relationships or interactions mentioned in the first interview and asked interviewees to define key terms (empathy, compassion, and engagement) and describe how these terms applied to the scenarios under discussion. Interviews were audio-recorded, transcribed using the web-based platform otter.ai, and then reviewed by the interviewer to correct any mistakes. A constructivist grounded theory approach (Mills et al., 2006) was used to identify themes and affiliated codes; two researchers independently coded interviews and compared results to identify inconsistencies, which were then resolved through discussion. Codes were thematically grouped to identify mediating factors in the model for how empathy is conceptualized. Below we present results from two of the participants interviewed, Sean and Jacob (pseudonyms).

## Results

Contextual information, encompassing personal and environmental details, aids faculty members in noticing and evaluating situations, deciding on actions, and mediates between *Attention/Noticing* and *Cognitive Empathy*. Sean explicitly discusses how they notice their graduate student struggling, reach out for information, and use that to develop an understanding of the student's personal and affective struggles:

> Milestones were getting missed… and I would ask…they'd honestly be pretty upfront and say. I'm struggling and going through X, Y, and Z, so it was a conversation. It's usually a pattern, as a faculty, as you gain experience with how students are and so on..you sense patterns. In this particular case there's a lot of things; one of the things is an imposter syndrome thing that's going on.

Similarly, Jacob shows a similar process of dialog leading to awareness of important contextual information when discussing an undergraduate learning assistant missing shifts:

> They'd show up or just not at all without letting me know and that kept happening throughout the semester and that was difficult to deal with. Because when they were present, they were excellent. But they were not working with me consistently. They just weren't there too often. And I talked with them about it. And it wasn't that they were blowing things off on purpose. Like they had serious struggles in their life that were manifesting as mental emotional health problems too. And so they weren't able to consistently show up and be what I needed them to be.

A struggle to communicate, which we label *dis-communication*, impedes the empathetic pathways, most evidently in the stage between Cognitive Empathy and Empathetic Action. Sean expresses his frustrations with communication as an artifact of external factors and its impact on empathetic action, saying:

> It is disheartening, maybe is the word to me, that people suffer from this and that this person is struggling with that in particular…Certain things having to do with like, the culture of science and things like that, like, you know, those are things we can communicate about and figure out and try to isolate, okay, like, why is this an impediment to you, and how can we get past it and what are the tools we could use, you know, things like that. And so, yeah, that was disheartening. And I think, yeah, it's just in general, it's frustrating to see somebody who can do it. And it's like, I know you can do it. I've seen you do it. You're doing it right now. Right? [And they] say to you, I can't do it.

## Discussion

Sean and Jacob both show the importance of context. Sean develops empathy for the student as they work to understand what the student is going through, a process made possible by the contextual information the student communicates. We see evidence of this context in Sean naming a phenomenon --- imposter syndrome --- and connecting this to a pattern they observe in students. We note that Sean does not describe personal experiences with imposter syndrome, suggesting this is a cognitive empathy. This is reinforced in Sean's frustration, which arises from their inability to effectively communicate their own positive estimation to the student. It is "disheartening" that they cannot get past the culture of science, and "frustrating" that they cannot effectively communicate their belief that the student "can do it." Sean experiences emotions, but they are connected to the situation and its resolution, not a connection with the student that would indicate emotional empathy. Similarly, Jacob notices that their student is not showing up to their work shifts but develops cognitive empathy from contextual information about why. They say 'it wasn't like they were blowing things off on purpose' and they had some 'serious issues', showing an understanding that comes from the contextual information.

Sean's experience also illustrates a misalignment between intent and outcome, leading us to introduce the term *dis-communication*. This differs from miscommunication, as it goes beyond clarity to consider an individual's choices regarding what they want to know, questions to ask, or issues to address concurrently. It can hinder the gathering of contextual information or recognition of shared experiences and impede the translation of empathy into action. Dis-communication introduces factors such as power dynamics, expectations, and personal beliefs into the communication process and includes instances of unclear language, misaligned timelines, and missed opportunities. In essence, dis-communication functions as an inhibitor, affecting the links between attention and empathy development, as well as between cognitive empathy and empathetic action. In Sean's case, they are attempting to communicate their belief in the student's efficacy, but the student does not believe it. I.e., the information they are relaying is not conveying the intended meaning to the recipient and thus is not effectively communicated. This example extends past miscommunication as it highlights the interaction between individuals' internal states and the words through which messages are transmitted. When faculty do not notice racial, sexist, or ableist issues, this inhibits their communication with their students, an example of dis-communication. Students pick up insights about faculty based on what they choose to communicate. Therefore, when faculty members do not actively address or discuss certain topics, their silence itself conveys a message, regardless of intention. It should be noted that this paper is limited by only highlighting the experiences of two of our participants and thus our findings are not meant to be generalizable. Additionally, due to the scope of this study, we were unable to explicitly analyze faculty's awareness of systems of oppression within physics and the potential role of empathy in addressing them. Further work is needed to understand the cyclic nature between developing empathy and learning to pay more attention to oppressive actions.

## Conclusions

We have presented interview data that sheds light on how contextual information mediates cognitive empathy from the faculty perspective, further developing the established model of empathetic pathways. We illuminate how empathy development does not happen within an isolated individual but requires communication between the two participants. The path between noticing, developing empathy, and taking action is not linear, with a complicated relationship mediated by communication and subject to interference from dis-communication. Our study raises additional questions about effort and intention behind cognitive empathy and, in particular, what motivates individuals to engage in alternate perspective-taking and empathetic action.

## Acknowledgments

This research was sponsored through NSF Awards #2149957, and # 2222337. We appreciate the collaborative efforts and insights provided by the members of CASTLE. Many thanks to RIT Capstone & summer REU programs.